\documentclass[11pt]{article}

\usepackage{amsmath}
\usepackage{amsfonts}
\usepackage{amssymb}
\usepackage{graphicx}
\usepackage{supertabular}
\usepackage{setspace}
\usepackage{xcolor}
\usepackage{mathptmx}

\RequirePackage{cite}%
\renewcommand{\citeleft}{\bgroup\normalfont[}%
\renewcommand{\citeright}{]\egroup}%

\newcommand{\n}{\noindent}
\newcommand{\nn}{\nonumber}

\newcommand{\be}{\begin{equation}}
\newcommand{\ee}{\end{equation}}
\newcommand{\ba}{\begin{eqnarray}}
\newcommand{\ea}{\end{eqnarray}}
\newcommand{\bal}{\begin{align}}
\newcommand{\eal}{\end{align}}

\newcommand{\dd}{{\rm d}}

\begin{document}

\title{\leftline{\footnotesize doi:10.1088/0264-9381/28/14/148001}
\leftline{\footnotesize M. Azreg-A\"{\i}nou, \textit{Class. Quantum Grav.} \textbf{28} (2011) 148001 (2pp)}
\vskip0.5cm
Comment on `Spinning loop black holes'}
\author{Mustapha Azreg-A\"{\i}nou
\\Ba\c{s}kent University, Department of Mathematics, Ba\u{g}l\i ca Campus, Ankara, Turkey}
\date{}

\maketitle

\begin{abstract}
We review the derivations and conclusions made in Caravelli and Modesto (2010 \textit{Class. Quantum Grav.} \textbf{27} 245022) and show that most of the analysis performed there is not valid.

\vspace{3mm}

\n {\footnotesize\textbf{PACS numbers:} 04.60.Pp, 04.70.--s}

%\vspace{-4mm} \n \line(1,0){431} % for \onehalfspacing
\vspace{-3mm} \n \line(1,0){431} % for no spacing
\end{abstract}

%\medskip

In subsection 4.2 of~\cite{1}, the authors, upon performing a type 2 complexification on a spherically symmetric metric in retarded Eddington-Finkelstein (EF) coordinates, arrived at
\begin{align}
\dd s^2 = & G\dd u^2 + 2\dd u\dd r - \Sigma \dd \theta - 2a\sin^2\theta \dd r \dd \phi \nn\\
\label{c1}\quad & +[a^2(G-2)\sin^2\theta - \Sigma]\sin^2\theta \dd \phi^2 + 2a(1-G)\sin^2\theta \dd \phi \dd u\,,
\end{align}
where we have added the term $2\dd u\dd r$ missing in Eq.~(22) of~\cite{1}. $G$ and $\Sigma$ are functions of $(r,\theta)$
\begin{align}
\label{c2}& \Sigma(r,\theta)= \rho^2 + \frac{a_0^2}{\rho^2}\,,\quad G(r,\theta)=\frac{\rho^2-2mr}{\Sigma}=\frac{\Delta-a^2\sin^2\theta}{\Sigma}\\
& \text{with }\rho^2=r^2+a^2\cos^2\theta\,,\quad\Delta=r^2-2mr+a^2\,.
\end{align}

A claim was made in the first paragraph of page 14 of~\cite{1} that\footnote{In that paragraph the expression `$\Sigma(r,\theta)$ defined in (21)' should read `$\Sigma(r,\theta)$ defined in (44)'.} metric~\eqref{c1} can be written in the Boyer-Lindquist (BL) coordinates (where the metric coefficients of the mixed terms $\dd t\dd r$ and $\dd \phi\dd r$ are zero) and be brought to Eq.~(26) of~\cite{1}, with $\Sigma$ as given in~\eqref{c2}, by a `coordinate transformation' of the form
\begin{equation}\label{c3}
    \dd u=\dd t+g\dd r\,,\quad \dd \phi=\dd \phi'+h\dd r\,,
\end{equation}
where $g$ and $h$ are supposed to depend only on $r$. We show that this is not possible and that all the analysis and conclusions made in~\cite{1}, which are based on this claim, are not valid.

If we substitute~\eqref{c3} into~\eqref{c1}, then the requirement that $g_{tr}=0$ and $g_{\phi'r}=0$ leads to
\begin{equation}\label{c4}
    h(r)=-\frac{a}{\Sigma G+a^2\sin^2\theta}=-\frac{a}{\Delta(r)}\,,\quad
    g(r,\theta)=-\frac{\Sigma+a^2\sin^2\theta}{\Sigma G+a^2\sin^2\theta}=-\frac{r^2+a^2}{\Delta}-\frac{a_0^2}{\Delta \rho^2}\,.
\end{equation}
Since $g$ depends on $\theta$ through $\rho^2$, the system of equations~\eqref{c3} does not constitute a coordinate transformation. Thus, Eq.~(26) of~\cite{1}, with $\Sigma$ as given in~\eqref{c2}, does not describe the geometry of solution~\eqref{c1} in BL coordinates. Figures 8 and 9 of~\cite{1} plotting the Ricci scalar, which have been sketched using Eq.~(26) of~\cite{1} with $\Sigma$ defined by~\eqref{c2}, Eq.~(45) and the last paragraph of page 14 of~\cite{1} are not valid. Figures 10, 11 and 12 of~\cite{1} are likely not valid too.

We have derived an expression for the Ricci scalar of the solution using metric~\eqref{c1}
\begin{equation}\label{c5}
    R(r,\theta)=\frac{2 a_0^4 [\rho^4+(r^2+2mr) \rho^2-6 m r^3]}{\rho^2 (\rho^4+a_0^2)^3}+ \frac{4 a_0^2 \rho^2 [(4 a^2+3 r^2-3 a^2 \cos^2\theta) \rho^2-6 m r^3]}{(\rho^4+a_0^2)^3}\,,
\end{equation}
and found it different from that plotted in figures 8 and 9 of~\cite{1}. In the limit $r\to 0$ and $\theta \to \pi/2$, we have $\lim_{(r,\theta)\to (0,\pi/2)}R(r,\theta)=0$, which is not equal the limit given in the caption of figure 9 of~\cite{1}. A plot of the Ricci scalar~\eqref{c5}, as shown in figure 1, is manifestly different from that of figure 9 of~\cite{1}.
\begin{figure}[h]
\centering
  \includegraphics[width=0.5\textwidth]{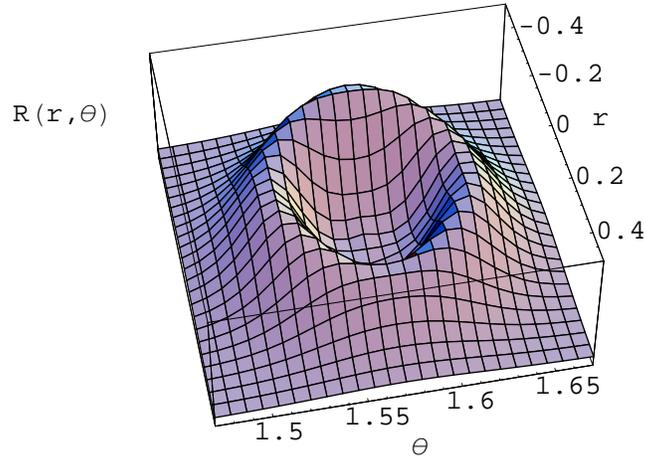}\\
  \caption{\footnotesize{The Ricci scalar $R(r,\theta)$ of metric~\eqref{c1} for $m=10$, $a=5$ and $a_0=0.1$.}}\label{Fig1}
\end{figure}

Similarly, the metric given in Eqs. (50) and (51) of~\cite{1} cannot be converted to BL form by transformation~\eqref{c3}, since this would require a dependence on $\theta$ of both functions $h$ and $g$. Thus, Eq. (55) of~\cite{1} and the claim made in the paragraph preceding it, as well as any conclusion based on Eq. (55) of~\cite{1}, are not valid. Our conclusion extends most likely to Eq. (72) of~\cite{1}, which has been derived using~\eqref{c3} as a `coordinate transformation' when $h$ and/or $g$ depend(s) on $\theta$.

\end{document}